\documentclass[prd,aps]{revtex4}

\usepackage{amsmath}
\usepackage{graphicx}
\usepackage{amssymb}
\usepackage{color}
\usepackage{bbold}

\newcommand{\be}{\begin{eqnarray}}
\newcommand{\ee}{\end{eqnarray}}
\newcommand{\bi}{\begin{itemize}}
\newcommand{\ei}{\end{itemize}}

\def\H{H}

\def\d{{\rm d}}

\def\1{{(1)}}
\def\2{{(2)}}
\def\3{{(3)}}

\def\vk{{\vec{k}}}
\def\vp{{\vec{p}}}

\def\vk{\vec{k}}

\def\vr{\vec{r}}

\newcounter{hran}

\def\MSbar{\relax\ifmmode\overline{\rm MS}\else{$\overline{\rm MS}${ }}\fi}

\def\H{H}

\def\d{{\rm d}}

\def\1{{(1)}}
\def\2{{(2)}}
\def\3{{(3)}}

\def\vk{{\vec{k}}}

\def\vk{\vec{k}}

\numberwithin{equation}{section}
\begin{document}

\def\thefootnote{\fnsymbol{footnote}}










\renewcommand{\topfraction}{0.99}
\renewcommand{\bottomfraction}{0.99}

\title{ \Large\bf On the Halo Velocity Bias}

\vskip 2cm

\author{Matteo Biagetti$^{(1)}$, Vincent Desjacques$^{(1)}$, Alex Kehagias$^{(1,2)}$ and Antonio Riotto$^{(1)}$}

\vskip 5cm
\address{$^{(1)}$Universit\'e de Gen\`eve, Department of Theoretical Physics and Center for Astroparticle Physics (CAP),\\ 24 quai E. Ansermet, CH-1211 Geneva 4, Switzerland}

\address{$^{(2)}$Physics Division, National Technical University of Athens, 15780 Zografou Campus, 
Athens, Greece}

\date{\today}

\begin{abstract}
\noindent
It has been recently shown that  any halo velocity bias present in the initial conditions does not 
decay to unity, in agreement with predictions from peak theory. However, this is at odds with 
the standard formalism based on the coupled-fluids approximation for the coevolution of dark matter 
and halos.
Starting from conservation laws in phase space, we discuss why the  fluid  momentum conservation equation 
for the biased tracers needs to be modified in accordance with the change advocated in Baldauf, Desjacques 
\& Seljak (2014).
Our findings indicate that a correct description of the halo properties should properly take into account 
peak constraints when starting from the Vlasov-Boltzmann equation.
\end{abstract}

\def\thefootnote{\arabic{footnote}}
\setcounter{footnote}{0}
\maketitle



\section{Introduction}\pagenumbering{arabic}
\noindent
While the existence of a spatial bias between the halos hosting galaxies and the underlying 
dark matter (DM) distribution has been firmly established since the pioneering work of 
\cite{kaiser}, the presence of a halo velocity bias is still being debated. 
Clearly, whereas galaxy velocities can be physically biased ({\it i.e.} on an object-by-object 
basis) owing to differences between the dark matter and baryon velocity fields (see, {\it e.g.}, 
Refs. \cite{tseliak,yoo}), by construction halo locally flow with the dark matter (in Einstein's
theory of gravity), so that a halo velocity bias can only arise statistically. Namely, it must 
be the statistical manifestation of a selection effect, which should naturally arise since 
virialized structures preferentially trace overdense regions of the Universe
\cite{SC,PS}.

A step forward in the understanding of halo bias has been recently made in Ref. \cite{vbias} 
where, through N-body simulations, the authors have measured a halo velocity bias $b_v(k)$, 

\be
\label{biasv}
v_{\rm h}(\vk,t)=b_v(k)\,v_{\rm dm}(\vk,t), \,\,\,\,b_v(k)=(1-R_v^2k^2),
\ee
which appears to remain constant throughout the cosmic  time $t$ until virialization. Here, 
$R_v$ is the typical scale of the halo velocity bias

\be
R_v^2=\frac{\sigma_0^2}{\sigma_1^2},\,\,\,\,\sigma_j^2=\int\frac{\d^3 k}{(2\pi)^3}\, k^{2j} 
P_{\rm dm}(k)W^2(KR),
\ee 
$P_{\rm dm}(k)$ is the DM power spectrum and $W(x)$  is a spherically symmetric smoothing 
kernel. This statistical effect is consistent with the relative suppression of the halo 
velocity divergence power spectrum at late time subsequently reported in Ref.  \cite{jennings}, 
although this type of measurement is more prone to systematics arising from sparse sampling
(see {\it e.g.} Refs.  \cite{percival,eliapk,zhang}).

Whereas the finding of  Ref. \cite{vbias} is in full agreement with the  peak model,  which indeed
predicts the existence of a linear, statistical halo velocity bias which remains constant 
with time \cite{des08,p1,p2}, it seems at odds with the prediction based on the coupled-fluids 
approximation for the coevolution of DM and halos \cite{nonlocalbias}.
The latter is widely used to compute the time evolution of bias \cite{f1,f2,eliapt} and is based 
on the idea of following the evolution over cosmic time and in Eulerian space of the halo 
progenitors - the so-called proto-halos - until their virialization.
While  their shapes and topology change as a function of time (smaller substructures gradually 
merge to form the final halo), their centre of mass moves along a well-defined trajectory 
determined by the surrounding mass density field.
Therefore, unlike virialized halos that experience merging, by construction proto-halos
always preserve their identity. 
Their total number is therefore conserved over time, and one 
can write a continuity equation and an Euler equation for their number density and velocity, 
respectively. Nevertheless, this approach predicts that any Eulerian  velocity bias rapidly 
decays to unity \cite{nonlocalbias}

\be
\label{gh}
b^{\rm E}_v(k,t)=1+D^{-3/2}(t)(b_v(k)-1),
\ee
where $D(t)$ is the  linear growth rate normalized to unity at the collapse redshift. 
To reconcile these two apparently contradictory results, the authors of Ref. \cite{vbias} 
argued that the Euler equation for halos should be changed from

\be
\label{halolim}
\dot{\theta}_{\rm h}+H\theta _{\rm h}+\frac{3}{2}H^2\Omega_{\rm dm}\delta+\cdots=0,\,\,\,\,
{\theta}_{\rm h}=\vec{\nabla}\cdot \vec{v}_{\rm h},
\ee
which predicts the incorrect behavior (\ref{gh}), to 

\be
\label{true}
\dot{\theta}_{\rm h}+H\theta _{\rm h}+\frac{3}{2}b_v(k)H^2\Omega_{\rm dm}\delta_{\rm dm}+\cdots=0,\,\,\,\,
{\theta}_{\rm h}=\vec{\nabla}\cdot \vec{v}_{\rm h},
\ee
where  $H$ is the Hubble rate and $\Omega_{\rm dm}$ parametrizes the abundance of DM and 
the dots stand for higher-order terms. 
Their physical interpretation is that the gravitational force acting on DM halos is 
statistically biased. Together with the Euler equation for DM

\be
\label{correct}
\dot{\theta}_{\rm dm}+H\theta_{\rm dm} +\frac{3}{2}H^2\Omega_{\rm dm}\delta_{\rm dm}+\cdots=0, \,\,\,\,
{\theta}_{\rm dm}=\vec{\nabla}\cdot \vec{v}_{\rm dm},
\ee
one indeed recovers the behavior (\ref{biasv}) and the Eulerian velocity halo bias does not decay 
in time to unity.
 
Another reason why Eq. (\ref{halolim}) cannot describe the momentum evolution of halos is the
fact that it does not differ at all from (\ref{correct}). Consequently, it is not possible that 
Eq. (\ref{halolim}) describes clustered objects like halos (or peaks), which are different from 
the smooth  DM distribution as they are supposed to be located at points where 
$\vec{\nabla}\delta_{\rm dm}=\vec{0}$ and where the smoothed density contrast is larger than some 
value $\nu\sigma_0$, being $\nu$ the peak height. 
In other words, Eq. (\ref{halolim}) does not contain any information about 
the fact we are dealing with peaks.
 
The goal of this short note is to explain -- at a somewhat more fundamental level than done in 
\cite{vbias} -- why and how one needs to modify the momentum fluid equation  for the coupled-fluids 
of halos and DM to obtain Eqs. (\ref{true}) and (\ref{correct}) as advocated in \cite{vbias}. The paper is organized as follows. In section II we present a derivation of the fluid equation from first principles and deal with the effective force felt by peaks in section III. Finally, section IV contains our conclusions and further comments.
 
 \section{From the  Klimontovich-Dupree equation to the fluid equation for halos}\label{sec:nonfluid}
\noindent
Let us start from the single particle phase space density
\be
f_{\rm K}(\vr,\vp,t)=\sum_i\delta_{\rm D}\left[\vr-\vr_i(t)\right]\delta_{\rm D}\left[\vp-\vp_i(t)\right],
\ee
where ``K'' indicate the so-called Klimontovich density \cite{klim,bert96}. We are following the phase space trajectories of single DM ``particles'' without any averaging, instead we are considering only one realization of a universe. We have used the cosmic time $t$ and halos will be eventually  identified at the  points where maxima of the DM density contrast are located and with a smoothed
density contrast larger than some $\nu\sigma_0$.  
The Klimontovich density obeys the Klimontovich-Dupree equation

\be
\label{kd}
\frac{\partial f_{\rm K}}{\partial t}+\vec{p}\cdot \frac{\partial}{\partial\vr}f_{\rm K}
-\vec{\nabla}\Phi_{\rm K}\cdot\frac{\partial}{\partial\vp} f_{\rm K}=0,
\ee
where 

\be
\vec{\nabla}\Phi_{\rm K}=G_{\rm N}\int\d^3r'\d^3p'f_{\rm K}(\vr',\vp',t)\frac{(\vr-\vr')}{\left|\vr-\vr'\right|^3}.
\ee
The Klimontovich density following the trajectories of all the single particles, some coarse graining is needed in order to handle the huge amount of information encoded in this quantity. Macrostates can be identified by averaging in a standard way over a statistical ensemble of microstates with similar phase space density in small volumes containing a sufficient amount of particles.

Denoting this averaging using angle brackets $\langle\cdots\rangle$, the first of the distribution functions
is given by

\be
\Big<f_{\rm K}(\vr,\vp,t)\Big>=
\Big<\sum_i\delta_{\rm D}\left[\vr-\vr_i(t)\right]\delta_{\rm D}\left[\vp-\vp_i(t)\right]\Big>=f(\vr,\vp,t).
\ee
Therefore, on averaging the Klimontovich equation (\ref{kd}) over the ensemble of realizations, we obtain 

\be
\label{poi}
\frac{\partial f}{\partial t}+\vec{p}\cdot \frac{\partial f}{\partial\vr}
-\langle\vec{\nabla}\Phi\rangle\cdot\frac{\partial f}{\partial\vp} + \cdots=0,
\ee
where

\be
\label{dd}
\langle\vec{\nabla}\Phi\rangle=
G_{\rm N}\int\d^3r'\d^3p'f(\vr',\vp',t)\frac{(\vr-\vr')}{\left|\vr-\vr'\right|^3}
\nonumber \;,
\ee
and the $\cdots$ stand for  terms arising at higher-order in perturbation theory when  correlations
introduced by gravitational clustering are taken into account. We will comment on this point later on.

The derivation above highlights that one should compute the averaged (in a statistical sense) gravitational force (per unit mass) $\langle\vec{\nabla}\Phi\rangle$. 
The smooth DM distribution is described by the Vlasov equation, such that the average force $\langle\vec{\nabla}\Phi\rangle$ can be replaced by the Poisson equation

\be
\nabla^2 \Phi=\frac{3}{2}H^2\delta_{\rm dm}.
\ee
However, in the particular case of density peaks, the average force exerted by the smooth DM component is statistically biased owing to the fact that peaks stream towards (or move apart from) each other {\it more coherently} in high (low) density environments. 
One should remember though that  on an object-by-object basis, all the particle species (DM, halos etc.) experience the same force, in agreement with Einstein's equivalence principle.

In order to use Eq. (\ref{poi}) to describe the halo phase-space  it is necessary therefore  to consider the mean shift in 
$\langle\vec{\nabla}\Phi\rangle$ in the vicinity of a peak. Since $\vec{\nabla}\Phi$ correlates only with $\vec{\nabla}\delta_{\rm dm}$ (at linear order), one has to compute the conditional probability of having at a given peak location a given value of $\vec{\nabla}\Phi$
given the fact that at the same point the gradient of the DM density contrast has a given value $\vec{\nabla}\delta_{\rm dm}$. We will perform this calculation in the next  section.
%
%
%
%
%
%

\section{The effective force for peaks}
\noindent
In order to understand the change in the force felt by the peaks on average, we simply have to calculate the average force (\ref{dd}) subject to the peak constraint, that is, $\vec\nabla\delta_{\rm dm}=\vec{0}$ and the corresponding Hessian is negative  (since peaks are identified with local maxima of the DM over density field). 

 We will restrict ourselves at the 
linear level in perturbation theory in such a way that,  at early times, the DM density $\delta_{\rm dm}$ and  $\vec\nabla\Phi$  are Gaussian variables.  We can therefore apply the theorem presented, for instance,  in Refs. \cite{bbks,bm} stating that the conditional probability of zero-mean Gaussian variables $Y_A$ and $Y_B$ is itself a Gaussian variable with mean

\be
\Big<Y_B\Big|Y_A\Big> \equiv \frac{\Big<Y_B \otimes Y_A\Big>}{\Big<Y_A \otimes Y_A \Big>}Y_A
\ee
and covariance matrix

\be
{\rm C}(Y_B,Y_A) \equiv \Big<Y_B \otimes Y_B \Big> - \frac{\Big<Y_B \otimes Y_A\Big>}{\Big<Y_A \otimes Y_A \Big>}\Big<Y_A \otimes Y_B \Big>.
\ee
Let us identify $Y_A$ with $\vec\nabla\delta_{\rm dm}$ and $Y_B$ with $\vec\nabla\Phi$. 
Therefore,  the mean shift in 
$\langle\vec{\nabla}\Phi\rangle$ in the vicinity of a peak is given by 

\begin{eqnarray}
\Big<\vec{\nabla}\Phi\Big |\vec{\nabla}\delta_{\rm dm}\Big>&=&\frac{\Big<\vec{\nabla}\Phi\vec{\nabla}\delta_{\rm dm}\Big>}{\Big<\left(\vec{\nabla}\delta_{\rm dm}\right)^2\Big>}\,\vec{\nabla}\delta_{\rm dm}\nonumber\\
&=&-\frac{3}{2}H^2
\frac{\Big<\delta^2_{\rm dm}\Big>}{\Big<\left(\vec{\nabla}\delta_{\rm dm}\right)^2\Big>}\,\vec{\nabla}\delta_{\rm dm}\nonumber\\
&=&
-\frac{3}{2}H^2\,\frac{\sigma_0^2}{\sigma_1^2}\,\vec{\nabla}\delta_{\rm dm}\nonumber\\
&=&-\frac{3}{2}H^2\,R_v^2\,\vec{\nabla}\delta_{\rm dm}.
\end{eqnarray}
We reiterate that, in this expression, we have assumed that both $\vec{\nabla}\Phi$ and $\vec{\nabla}\delta_{\rm dm}$ are Gaussian-distributed and therefore the result is valid only at the linear level. At higher-order in perturbation theory a modification should be expected when going to smaller distances. 

We now proceed to compute the  covariance matrix for $\vec{\nabla}\Phi$ given the constraint. First, we provide the various elements

\begin{eqnarray}
\Big<\vec{\nabla}\Phi\otimes\vec{\nabla}\Phi\Big>&=&\frac{\alpha^2}{3}\sigma^2_{-1}\,
\mbox{\Large \(  \mathbb{1} \)}_{3\times 3},
\nonumber\\
\Big<\vec{\nabla}\Phi\otimes\vec{\nabla}\delta_{\rm dm}\Big>&=&-\frac{\alpha}{3}\sigma^2_{0}\,\mbox{\Large\(  \mathbb{1} \)}_{3\times 3},\nonumber\\
\Big<\vec{\nabla}\delta_{\rm dm}\otimes\vec{\nabla}\Phi\Big>&=&-\frac{\alpha}{3}\sigma^2_{0}\,\mbox{\Large\(  \mathbb{1} \)}_{3\times 3},\nonumber\\
\Big<\vec{\nabla}\delta_{\rm dm}\otimes\vec{\nabla}\delta_{\rm dm}\Big>&=&\frac{1}{3}\sigma^2_{1}\,\mbox{\Large\(  \mathbb{1} \)}_{3\times 3},
\end{eqnarray}
where we have defined $\alpha=3\H^2\Omega_{\rm dm}/2$. The covariance matrix for $\vec{\nabla}\Phi$ given the constraint is therefore given by

\be
\label{cov}
C\left(\vec{\nabla}\Phi,\vec{\nabla}\Phi\right)=\frac{\alpha^2}{3}\left(\sigma^2_{-1}-\frac{\sigma^4_{0}}{\sigma^2_{1}}\right)\,\mbox{\Large\(  \mathbb{1} \)}_{3\times 3},
\ee
a result first derived in Ref.  \cite{bbks}.
Even though at the location of the peak  the mean shift of the gravitational force vanishes and, therefore, at the peak-by-peak level there is no extra force, its variance is not simply proportional to $\sigma^2_{-1}$, but it receives  a correction when the statistical ensemble average is taken. This effect can be simply captured by replacing the  force $ \vec{\nabla}\Phi$ experienced by the peaks by

\be
 \vec{\nabla}\Phi_{\rm eff}=\vec{\nabla}\Phi+\frac{3}{2}H^2\Omega_{\rm dm}\,R_v^2\,\vec{\nabla}\delta_{\rm dm},
\ee
which is precisely the velocity bias relation proposed in \cite{des08,p1,p2} since the force is proportional to the velocity in linear theory.
Indeed, the variance of this shifted quantity has zero mean and variance given by 

\be
\Big<\left( \vec{\nabla}\Phi_{\rm eff}\right)^2\Big>=\frac{\alpha^2}{3}\left(\sigma^2_{-1}-\frac{\sigma^4_{0}}{\sigma^2_{1}}\right),
\ee
which coincides with Eq. (\ref{cov}). Inserting this result into Eq. (\ref{poi}) for halos

\be
\label{poii}
\frac{\partial f_{\rm h}}{\partial t}+\vec{p}\cdot \frac{\partial f_{\rm h}}{\partial\vr}
-\vec{\nabla}\Phi_{\rm eff}\cdot\frac{\partial f_{\rm h}}{\partial\vp} +\cdots=0,
\ee
multiplying now Eq. (\ref{poi}) by $\vp$, reinstating the expansion of the universe, integrating over the momenta and applying the gradient on both sides, we precisely get Eq. (\ref{true}). 
Let us stress again that the effective force felt by the halos is of pure statistical origin, the definition of $\Phi_{\rm eff}$ being deduced from the covariance rather then from the average of $\vec{\nabla}\Phi_{\rm eff}$, which is still vanishing (at linear order).

\section{Conclusions and further comments}
\noindent
In this short note, we have considered conservation laws in phase 
space to explain why the gravitational force acting on biased 
tracers of the large-scale structure is itself biased, as was already 
noted in Refs.  \cite{p2,vbias}.
Consequently, the halo momentum conservation equation is modified 
accordingly, in agreement with the change advocated in Ref. \cite{vbias}. 
This modification is responsible for the time constancy of the halo 
velocity bias (at the linear level).

We conclude by some remarks. Like in Refs. \cite{des08,p1,p2,vbias}, our 
derivation of the shift in the gravitational force felt by halos 
has made use of the properties of   Gaussian random fields, so it  
applies at early time and on sufficiently large scales, when 
linear perturbation theory holds. It would be interesting to 
understand  what happens when non-linearities become important. 
Along these lines, the work done in Ref. \cite{bm} might be useful. 
Furthermore, our derivation makes also clear that, when dealing 
with halos, the starting Vlasov-Boltzmann equation is highly 
non-trivial. Indeed, going to higher-order, not only the average 
force receives extra contribution, but also the equation (\ref{poii}) 
is modified into \cite{bert96}

\be
\frac{\partial f_{\rm h}}{\partial t}+\vec{p}\cdot \frac{\partial f_{\rm h}}{\partial\vr}
-\vec{\nabla}\Phi_{\rm eff}\cdot\frac{\partial f_{\rm h}}{\partial\vp} =\vec{\nabla}_{\vp}\cdot \vec{F}_{\rm h},
\ee
where

\be
\vec{F}_{\rm h}(\vr,\vp,t)={\rm Cov}\left[\vec\nabla\Phi_{\rm K}(\vr,t),f_{\rm K}(\vr,\vp,t)\right]= 
G_{\rm N}\int\d^3r'\, \d^3 \vp'\, \frac{(\vr-\vr')}{\left|\vr-\vr'\right|^3} f_{\rm 2c}(\vr,\vp,\vr',\vp',t),
\ee
and 

\be
f_{\rm 2c}(\vr,\vp,\vr',\vp',t)=f_{\rm 2}(\vr,\vp,\vr',\vp',t)-f(\vr,\vp,t)f(\vr',\vp',t)
\ee
is the irreducible   two-particle correlation function, which has to be computed with the necessary peak constraints.
The latter cannot be neglected because it may be much larger than the 
product of the single-particle terms  due to strong
gravitational clustering. The vector  $\vec{F}_{\rm h}$  represents a correlated
force density and arises from fluctuations in the ensemble averaged gravitational 
potential due to
clustering of the matter distribution. This force density must be computed in the 
presence of the peak constraint, and generally gives rise to drift forces and  
diffusion in velocity-space. 
Added up to the fact that an initial constraint on the halo shifts the mean density 
and causes the density and velocity to be correlated hence, this changes the halo 
phase space density. Any theory of bias should therefore account of these effects. 
Work along these lines is in progress \cite{us}.

 \section*{Acknowledgments}
\noindent
The research of A.K. was implemented under the Aristeia Action of the Operational Programme 
Education and Lifelong Learning and is co-funded by the European Social Fund (ESF) and National 
Resources. A.K. is also partially supported by European Union's Seventh Framework Programme 
(FP7/2007-2013) under REA grant agreement n. 329083. A.R. is supported by the Swiss National
Science Foundation (SNSF), project `The non-Gaussian Universe" (project number: 200021140236). 
M.B. and V.D. acknowledge support by the Swiss National Science Foundation.


\begin{small}

\end{small}


\begin{thebibliography}{99}  

\bibitem{kaiser} N. Kaiser,
  Astrophys. J. Lett. {\bf 284}, L9 (1984)

\bibitem{tseliak} D. Tseliakhovich, C.M. Hirata,
  Phys. Rev. D {\bf 82}, 083520 (2010), arXiv:1005.2416 [astro-ph.CO].

\bibitem{yoo} J. Yoo, U. N. Dalal, U. Seljak,
  JCAP {\bf 1107}, 018 (2011)
  [arXiv:1105.3732 [astro-ph.CO]].

\bibitem{SC} J.E. Gunn, J.R. Gott III,
  Astrophys. J. {\bf 176}, 1 (1972).

\bibitem{PS} W.H. Press, P. Schechter,
  Astrophys. J. {\bf 187}, 425 (1974).


\bibitem{vbias} T.~Baldauf, V.~Desjacques and U.~Seljak,
  arXiv:1405.5885 [astro-ph.CO].

\bibitem{jennings} E. Jennings, C.M. Baugh, D. Hatt,
  arXiv:1407.7296 [astro-ph.CO].

\bibitem{percival} W.J. Percival, M. White,
  Mon. Not. R. Astron. Soc. {\bf 393}, 297 (2009)
  arXiv:0808.0003 [astro-ph.CO].

\bibitem{eliapk} A. Elia, A.D. Ludlow, C. Porciani,
  Mon. Not. R. Astron. Soc. {\bf 421}, 3472 (2012)
  arXiv:1111.4211 [astro-ph.CO].

\bibitem{zhang} P. Zhang, Y. Zheng, Y. Jing,
  arXiv:1405.7125


\bibitem{des08} V.~Desjacques,
  Phys.\ Rev.\ D\ {\bf 78}, 103503 (2008)
  [arXiv:0806.0007 [astro-ph.CO]].

\bibitem{p1} V. Desjacques and R. K. Sheth, Phys. Rev. D {\bf 81}, 023526
(2010), arXiv:0909.4544 [astro-ph.CO].

\bibitem{p2} V. Desjacques, M. Crocce, R. Scoccimarro, and R. K.
Sheth, Phys. Rev. D {\bf 82}, 103529 (2010), arXiv:1009.3449
[astro-ph.CO].

\bibitem{nonlocalbias} K.~C.~Chan, R.~Scoccimarro and R.~K.~Sheth,
  Phys.\ Rev.\ D {\bf 85}, 083509 (2012)
  [arXiv:1201.3614 [astro-ph.CO]].

\bibitem{f1} J. N. Fry, Astrophys. J. {\bf 461}, L65 (1996).

\bibitem{f2} M. Tegmark and P. J. E. Peebles, Astrophys. J. Lett.
{\bf 500}, L79 (1998), astro-ph/9804067.

\bibitem{eliapt} A. Elia, S. Kulkarni, C. Porciani, M. Pietroni, and
S. Matarrese, Mon. Not. R. Astron. Soc. {\bf 416}, 1703
(2011), arXiv:1012.4833 [astro-ph.CO].

\bibitem{klim}
Yu. L. Klimontovich, The Statistical Theory of Non-Equilibrium Processes in a Plasma (MIT Press, Cambridge, 1967).

\bibitem{bert96} E. Bertschinger, in {\it Cosmology and Large Scale Structure}, proceedings of the  1996 Les Houches Summer School, 
Session LX, ed. R. Schaeffer, J. Silk, M. Spiro, and J. Zinn-Justin (Amsterdam: Elsevier
Science), 273.

\bibitem{bbks}
  J.~M.~Bardeen, J.~R.~Bond, N.~Kaiser and A.~S.~Szalay,
  Astrophys.\ J.\  {\bf 304} (1986) 15.

 \bibitem{bm} C.~-P.~Ma and E.~Bertschinger,
  Astrophys.\ J.\  {\bf 612}, 28 (2004)
  [astro-ph/0311049].
 



\bibitem{us} M. Biagetti, V. Desjaques, A. Kehagias and A. Riotto, to appear.











 \end{thebibliography}
\end{document}